# Sentiment Analysis for Education with R: packages, methods and practical applications

**Michelangelo Misuraca**
*University of Calabria*

**Alessia Forciniti, Germana Scepi, Maria Spano**
*University of Naples Federico II*

*Sentiment Analysis (SA) refers to a family of techniques at the crossroads of statistics, natural language processing, and computational linguistics. The primary goal is to detect the semantic orientation of individual opinions and comments expressed in written texts. There are several practical applications of SA in several domains. In an educational context, the use of this approach allows processing students' feedback, aiming at monitoring the teaching effectiveness of instructors and enhancing the learning experience. This paper wants to review the different R packages that can be used to carry on SA, comparing the implemented methods, discussing their characteristics, and showing how they perform by considering a simple example.*

## 1. Introduction

The incredible progress of computer technology and the growth of the Internet hastened the transition from analog to digital in several domains. In particular, the development of Web 2.0 produced a radical change in many aspects of everyday life. According to this new framework – developed over the past 15 years – users have the possibility of actively participating in the creation of information and data, sharing their contents on online platforms like social media or wikis. Because of the different nature of data that can be collected on online platforms – e.g., numbers, texts, emoticons – it is necessary to set up appropriate strategies to extract the most useful information and knowledge. This large and varied set of data cannot be processed manually. Nevertheless, automatic processing requires a significant computational effort. Traditional statistical techniques able to analyze quantitative and qualitative data can also be performed on textual data, pre-treating texts in a proper way, and structuring the resulting dataset. In recent years, considerable interest has been devoted to the analysis of user opinions. According to Pang and Lee (2008), *opinion mining* or *sentiment analysis* (SA) is a discipline at the crossroads of statistics, natural language processing, and computational linguistics, which tries to detect the opinions expressed in natural language texts. SA can be used to detect different emotional states and reveal a variety of behavioral and attitudinal patterns. These states or patterns, in general, can be detected through different techniques, such as process mining or discourse analysis. A large number of papers mention SA in the context of the so-called *polarity classification* (e.g.,

Cambria, Schuller, Xia, & Havasi, 2013). The main goal is to classify texts written in natural language, considering their semantic polarity and distinguishing positive and negative forms. It is difficult to extract useful information from opinions and then to understand, summarize, and organize this information into usable forms (Balahur & Jacquet, 2015). Due to the differences in the potential sources of opinions and their unique characteristics, there is not a "one-size-fits-all" approach (Petz, *et al.*, 2014).

The (r)evolution induced by new technologies also involved Education, allowing a significant improvement in teaching and learning activities as well as in the communication between instructors and students. The availability of large repositories of educational data led to the development of *learning analytics* (Verbert, *et al.*, 2013) and the formalization of a specialized branch of data mining known as *educational data mining* (Slater, *et al.*, 2017). These tools aim at providing meaningful insights, supporting the design of appropriate interventions for improving teaching practices and learning processes (Dietz-Uhler & Hurn, 2013). In particular, determining students' views by collecting and processing feedback on their learning experiences is widely recognized as a central strategy for assessing the quality of teaching at most educational institutions (Mandouit, 2016).

At the end of each academic term, students are commonly required to participate in post-course surveys to gather their experience. This process enables instructors and administrators to examine students' evaluations and then enhance the learning processes. The questionnaires include both close-ended questions and open-ended questions. The first ones are often in the form of Likert-scale questions and aim at capturing students' assessment in the form of numerical ratings. The second ones allow collecting written comments or suggestions that reflect the personal feelings and perceptions of the students. The majority of institutions paid great attention to quantitative feedback, easy to summarize and analyze with statistical techniques. The qualitative comments were instead not fully considered, even if they can provide valuable insights. Moreover, the use of social media and online platforms increased the amount of available textual content, making necessary to implement proper strategies of analysis. These channels allow real-time monitoring of students' opinions, mainly expressed in course forums or discussion groups, highlighting their mood, motivation, and understanding of topics. In this framework, SA can be used to highlight the semantic orientation of students' opinions, helping the evaluation process of the qualitative feedback.

In recent years, several papers showed the effectiveness of textual analyses and SA in an educational context, mainly referring to e-learning systems (e.g. Leong, Lee, & Mak, 2012; Aung & Myo, 2017; Jena, 2018) and *massive open online courses* (MOOCs: e.g. Wen, Yang, & Penstein Rosé, 2014; Moreno-Marcos, *et al.*, 2018). This paper wants to introduce educational researchers and practitioners to SA, showing how to calculate the polarities of a set of text with the most popular R packages designed for this task. In Section 2, we introduced some preliminary notation for textual data and the general background on polarity and semantic orientation. In Section 3, the capabilities of the five main R packages for SA are compared, with a brief note on other available resources. We then discussed the performances of the different packages in Sec-tion 4, highlighting both running times and effectiveness. In Section 5, we remarked some final comments on SA and the performed comparative study.

## 2. From words to sentiment: a textual approach

From a mathematical viewpoint, there are several ways to model a text written in natural language. In the framework of Text Mining, texts are usually coded as a *bag of words* (BoW). In this scheme, a text can be seen as an unordered collection of words, disregarding both grammatical relations and syntactic categories. Let consider a collection of $n$ texts $d_i$ ($i = 1, \ldots, n$). Text parsing allows identifying the different tokens used in the collection and the creation of a list of types, commonly known as vocabulary. A pre-processing step is often required for reducing the dimensionality of the vocabulary and eliminating the non-informative types (e.g., by using lists of stop-words). Since the expression "word" is too generic and does not encompass other useful combinations of words – like collocations and multi-words – in the following, *term* is used instead of word. According to the *vector space model* (Salton, Wong, & Yang, 1975), each $d_i$ is represented as a vector in the space spanned by the $p$ terms belonging to the vocabulary:

$$d_i = \{t_{i1}, \ldots, t_{im}, \ldots, t_{ip}\} \tag{1}$$

where $t_{im}$ is the importance of the $m$-th term in $d_i$. The document-vectors can be arranged in a matrix $X$ with $n$ rows (documents) and $p$ columns (terms). In a document, the importance of each term is usually measured by the *term frequency* – i.e., the number of occurrences of a term $m$ into a document $i$ – but other weighting schemes can also be considered. For a more comprehensive discussion on $t_{im}$ quantification, see Balbi & Misuraca (2005).

### *2.2 Polarity scores and semantic orientation*

The primary goal of SA is to classify documents by their *polarity* by identifying their *semantic orientation*. The expression polarity is commonly used in linguistics for distinguishing affirmative and negative terms (Löbner, 2000; Giannakidou, 2008). The overall evaluation of the polarity of the text gives the semantic orientation of the document. In the reference literature (e.g., Liu, 2015), three different levels of semantic orientation are considered:

1. the subjectivity/objectivity of a document (SO-orientation): the focus concerns if a text has a factual nature or instead expresses an opinion on its subjective matter;

2. the positivity/negativity of a document (PN-orientation): the focus concerns if a subjective text expresses a positive or negative opinion;

3. the positivity/negativity strength of a document (PN-strength): the focus concerns the identification of different grades of positive or negative sentiments in the text.

These three levels of semantic orientation are sequentially ordered by regarding the different granularity, but it is not mandatory to evaluate each one in an empirical research.

At the same time, we can consider three different levels for choosing an elementary unit of the analysis: a document level, a sentence-level, an aspect-level. The first two levels are usually entailed in the polarity-based SA, whereas the latter one is more used in a topic-based perspective. The document-level aims at defining the orientation of each text as a whole, i.e., if it expresses a positive or negative sentiment. In the sentence-level, each document is segmented into its sentences, and the polarity of each sentence is considered, then the polarity is synthesized at a document-level for evaluating the overall orientation (Tan, Na, Theng, & Chang, 2011).

The PN-orientation is evaluated by calculating a polarity score of -1, 0, and +1 for negative, neutral, and positive terms, respectively (Liu, Hu & Cheng, 2005). The polarity score of each term depends on the lexicon. A lexicon is a list of polarized terms. Both lexicons created manually (e.g., Tong, 2001) and lexicons created automatically or semi-automatically (e.g., Turney & Littman, 2003) can be considered. There are many papers in the literature dealing with the problem of choosing a proper lexicon (Bravo-Marquez, Mendoza, & Poblete, 2014). The available resources, even if developed for specific purposes, were used for several applicative domains.

Some authors proposed different scoring systems, defining the polarity not only in terms of the sign but also considering the strength of the positive/negative sentiment (Nielsen, 2011). In these cases, the lexicons consider a discrete or a continuous scale for the polarity, from the most negative to the most positive score. From the semantic orientation expressed as strength, it is ever possible to downgrade to a more straightforward evaluation as positivity/negativity.

### 2.3 Polarity scores and Sentiment Classification

Calculating the polarity scores of each document belonging to a given collection can be seen as the first step of a more sophisticated classification process, training a classifier with the different semantic orientations as categories. Nevertheless, it is also possible to consider the polarity detection has a lexicon-based approach to classification. When we aim at categorizing new documents considering the knowledge base obtained from the analysis of a document collection, a machine learning approach has to be considered. Different algorithms have been proposed in the literature for a document classification based on their semantic orientation. Several authors tried to systematically review the different approaches (e.g., Abirami & Gayathri, 2017). In a general way, it is possible to distinguish *probabilistic classifiers*, such as *naïve Bayes classifiers* (Kang, Yoo & Han, 2012) or *Bayesian networks* (Ortigosa-Hernández *et al.*, 2012); *linear classifiers*, such as *support vector machine* (Chin & Tseng, 2011); *decision trees* (Hu & Li, 2011). More recently, in the framework of linear classifiers and decision trees, considerable interest has been devoted to the *random forest* (e.g., Liu & Zhang, 2016) and the so-called *deep learning* (e.g., Dou, 2017).

## 3. R packages for Sentiment Analysis

R is a free software environment for statistical computing. It is widely used among quantitative researchers and data miners to perform various types of analysis. In the following, we de-scribe the main packages for SA. Table 1 lists the different packages, with the launch date and the total number of downloads until December 06th, 2019.

TABLE 1.
*Main R packages for Sentiment Analysis available at CRAN repositories.*

| Packages | Author(s) | Launch date | # of downloads |
|---|---|---|---|
| syuzhet | M. Jockers | 2015-02 | 343,840 |
| RSentiment | S. Bose | 2016-05 | 35,446 |
| sentimentr | T. Rinker | 2016-08 | 72,393 |
| SentimentAnalysis | S. Feuerriegel & N. Pröllochs | 2017-06 | 33,598 |
| meanr | D. Schmidt | 2017-06 | 7,907 |

We observed that the oldest packages had a higher number of downloads. Except for `syuzhet`, the first package designed explicitly for SA, `sentimentr` seemed to be the second most popular package. In the following, we showed the different approaches followed by the available packages, explaining the main functions and characteristics. Some other packages that can be used for SA are briefly described at the end of this section.

### 3.1 Data example

To show how the analyzed R packages perform sentiment analysis, we used as a case example the *Student Course Evaluation Comments* dataset built by Welch & Mihalcea (2017). This dataset is a collection of sentences extracted from a Facebook group where students can express their opinions about the courses and the instructors belonging to the Computer Science department at the University of Michigan. In the following, we referred to a subset of the original dataset available in the package `sentimentr` (Rinker, 2019), including only comments with an unambiguous polarity. This dataset, namely *course_evaluations*, has 566 units and two variables: the polarity scores (*sentiment*) and the students' comments (*text*). It includes 300 positive texts, 217 negative texts, and 49 neutral texts, with a score of +1, -1, and 0, respectively.

### 3.2 `syuzhet`

The `syuzhet` package (Jockers, 2017) allows calculating the polarity scores of a collection of documents by using different internal dictionaries. Particularly interesting is the possibility of accessing the robust sentiment extraction tool developed at Stanford University by the NLP group (Manning, *et al.*, 2014). The dictionaries used in `syuzhet` are: *syuzhet*, developed by the Nebraska Literary Lab under the direction of M. Jockers; *afinn*, developed by Nielsen (2011); *bing*, developed by Hu and Liu (2004); *nrc* developed by Turney and Mohammad (2010). The core function to calculate the polarity scores is `get_sentiment`:

```
get_sentiment(x, method="syuzhet", lexicon=NULL)
```

where `x` is the vector of strings, and `method` is the lexicon used for determining if a term is positive or negative. By default, the lexicon developed with the package is used. The other lexicons lead to slightly different results because each one uses a different scoring system at a term-level. The *bing* lexicon is a binary dictionary, the *afinn* and *syuzhet* lexicons are instead weighted dictionaries with a discrete and a continuous scale, respectively. The function calculates the polarity score of a text by summing the value associated with each term.

The following code provides an example on the *course_evaluations* dataset:

```r
1  ## Retrieving the "course_evaluations" dataset from sentimentr
2  library(sentimentr)
3  data(course_evaluations)
4  ## Extracting the text from the dataset as a vector of strings
5  x<-course_evaluations$text
6  ## Calculating sentiment scores
7  library(syuzhet)
8  s<-get_sentiment(x, method="syuzhet")
9  ## Calculating summary statistics on the sentiment distribution
10 summary(s)
```

```
> s
 [1]  0.50  0.50  0.75  0.80  0.00 -0.85  0.50  1.90  0.50  0.50 ...
> summary(s)
   Min. 1st Qu.  Median    Mean 3rd Qu.    Max.
-4.0000  0.0500  0.7500  0.6901  1.2500  3.9000
```

The package allows setting a custom lexicon in the polarity score calculation. This lexicon has to be imported as a data frame containing the terms and the corresponding polarity scores. Each score, at a term-level, can be both binary or weighted.

A `plot` function can be used to visualize the polarity score distribution of the document collection. By default, `syuzhet` considers the different texts as part of a single document, and the sentiment trajectory can be interpreted with respect to the overall *narrative time*. The function draws the values in a graph, shown in Figure 1, where the x-axis represents the time from the beginning to the end of the text, and the y-axis measures the degrees of the polarity scores:

```
plot(object, type="l", xlab="Narrative Time",
     ylab ="Emotional Valence", col="red")
           abline(h=0, col="black")
```

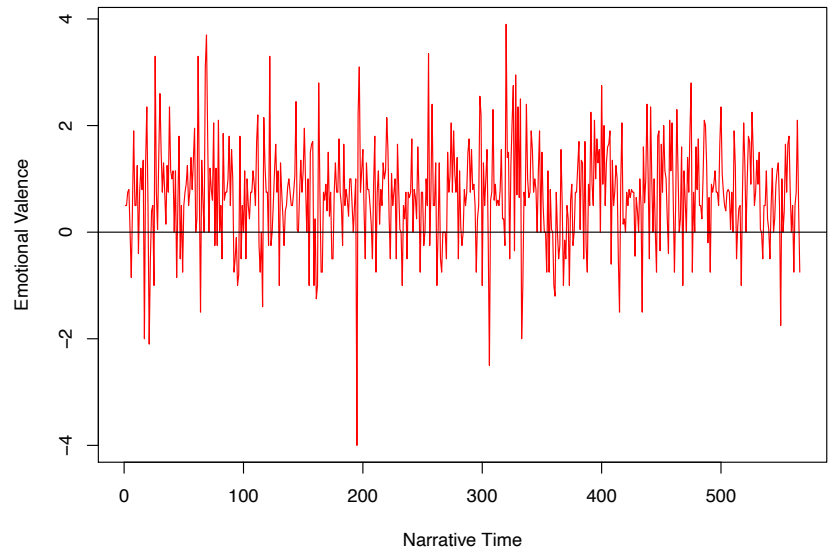

FIGURE 1. *Plot of the sentiment trajectory of the course_evaluation dataset from* `syuzhet`.

The package also allows comparing the shape of sentiment trajectories obtained from two or more different texts (Figure 2). The function `get_percentage_values` divides the collection in chunks with the same dimension, then calculates for each chunk the mean polarity score:

<div align="center">

`get_percentage_values(object, bins=100)`

</div>

where `object` is a vector with raw sentiment values, and `bins` are the number of splits.

```
1    ## Calculating the mean polarity scores per chunks of text
2    s_per<-getpercentage_values(s, bins=100)

> s_per
          1             2             3             4             5
  0.28333333    0.70833333    0.42000000    0.32500000    0.83333333  ...
```

It is important to note that combining texts into larger chunks dampen the extreme values of the emotional valence, so the mean values of longer passages tend to converge toward 0.

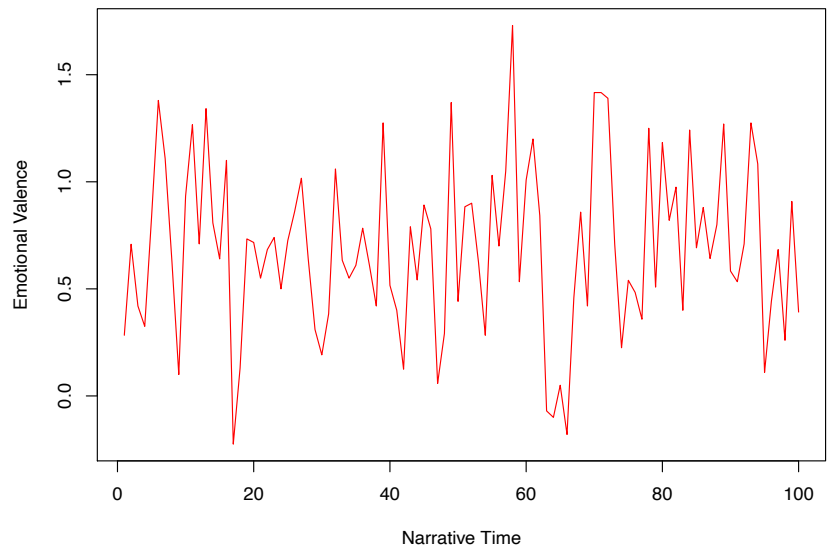

FIGURE 2. *Sentiment trajectory obtained from 100 chunks of the course_evaluation dataset.*

As an alternative, the package proposes a *discrete cosine transformation* (DCT) in combination with a low-pass filter[1], through the function `get_dct_transform`:

```
get_dct_transform(object, low_pass_size=5, x_reverse_len=100,
                  scale_vals=FALSE, scale_range=TRUE)
```

where `object` is a vector with raw sentiment values. This function passes scores with a frequency lower than a selected cutoff frequency and attenuates scores with frequencies higher than the cutoff frequency (Figure 3).

```
1  ## Performing Discrete Cosine transformation (DCT)
2  dct_values<-get_dct_transform(s, low_pass_size=5, x_reverse_len=100,
3  scale_vals=FALSE, scale_range=TRUE)

> dct_values
  [1]   0.9406102258   0.9328275559   0.9173263288   0.8942342028   ...
```

---

[1] A *low-pass filter* is a filter that passes signals with a frequency lower than a selected cutoff frequency and attenuates signals with frequencies higher than the cutoff frequency. The exact frequency response of the filter depends on the filter design.

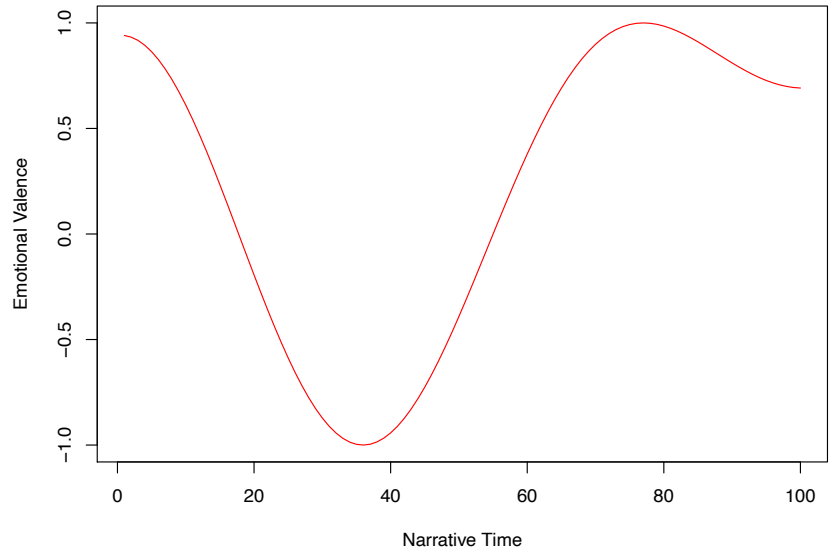

FIGURE 3. *Discrete cosine transformation of the sentiment trajectory.*

A peculiar task of `syuzhet` is calculating scores concerning emotions. The *nrc* dictionary is a list of English terms associated with a positive or negative polarity as well as to one of eight emotions (anger, fear, anticipation, trust, surprise, sadness, joy, and disgust). It is possible to select different languages. The underlying assumption is that, despite some cultural differences, the majority of emotional categories are stable across languages. The lexicon is available in more than one hundred languages, translating the original terms through Google Translate.

The emotional categorization can be obtained through the function `get_sentiment`, also specyfing the language into the parameters, or through the function `get_nrc_sentiment`:

$$get\_nrc\_sentiment(x, \ language="option")$$

where `x` is a vector of strings, and *option* is one of the available languages.

```
1   ## Detecting emotions on the texts
2   s_nrc<-get_nrc_sentiment(x)

> s_nrc[,1:8]
    anger anticipation disgust fear joy sadness surprise trust
1       0            0       0    0   1       0        0     1
2       0            0       0    0   0       0        0     0
3       0            0       0    0   1       0        0     1
...
```

It is also possible to visualize the distribution of emotions (Figure 4) by plotting the results:

```
barplot(sort(colSums(prop.table(s_nrc[,1:8]))), horiz=TRUE,
cex.names=0.7, las=1, col=palette("default"), xlab="Percentage")
```

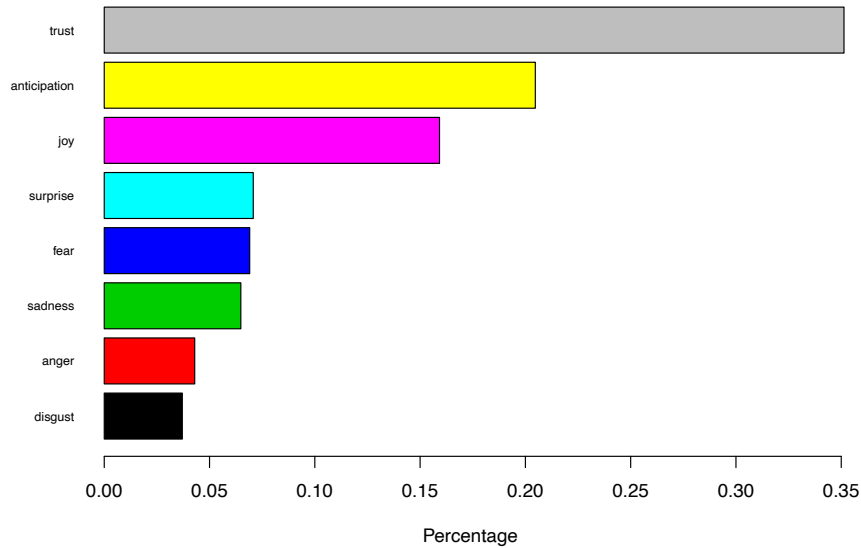

FIGURE 4. *Emotion percentage distribution of the course_evaluation dataset.*

### 3.3 `RSentiment`

The `RSentiment` package (Bose & Goswami, 2018) allows performing sentiment analysis on a sentence or a collection of sentences. The polarity scores are calculated by considering the grammatical role of each term, using a *Parts of Speech* (*POS*) tagging on the terms of the sentences. The primary function used in the package is `calculate_score`. After pre-processing the sentences – e.g., removing punctuations and normalizing the texts – the function checks for the POS of each term, the presence of negations such as "not" and "never", the presence of amplifiers such as "very" and "more". The dictionary used for determining if a term is positive or negative is the Hu and Liu lexicon. The overall score of a sentence is calculated as the difference between positive and negative terms. If there is an equal number of positive and negative terms, the sentence is considered neutral and is assigned a score equal to 0. Scores higher than 0 implies a positive sentiment of the sentences, whereas scores lower than 0 denote a negative sentiment. Interrogative sentences are marked with a score of 99 and considered sarcastic.

```
1  library(sentimentr)
2  data(course_evaluations)
3  library(RSentiment)
4  x<-course_evaluations$text
5  ## Calculating sentiment scores in RSentiment
6  s<-calculate_score(x)

[1] "Processing sentence: the labs in eecs 588 are not very engaging or
```

```
challenging in my opinion"
[1] "Processing sentence: i did not like eecs 417"
[1] "Processing sentence: my three favorite classes in eecs have been eecs
230 498 and 583"
...

> s
  [1]  0 -1  1  1  1 -1  1  5  1  0  0 -1  1  1 -1  2 -3  3  4  1 -3 -1
...
```

`RSentiment` gives the possibility of classifying the sentences into six sentiment categories: Very Negative, Negative, Neutral, Positive, Very Positive, and Sarcasm. This categorization is performed by using the function `calculate_sentiment`. This function assigns a tag to each value obtained as in the previous function, by subtracting the number of negative terms from the number of positive terms. Sentences with a score of -1 are marked as negative, sentences with a score lower than -1 are marked as very negative. In the same way, sentences with a score of +1 are marked as positive, sentences with a score higher than +1 are marked as very positive. The results are reported in a data frame listing the pre-processed text of each sentence and the corresponding polarity.

```
1  ## Calculating sentiment categories in RSentiment
2  s<-calculate_sentiment(x)

[1] "Processing sentence: the labs in eecs 588 are not very engaging or
challenging in my opinion"
[1] "Processing sentence: i did not like eecs 417"
[1] "Processing sentence: my three favorite classes in eecs have been eecs
230 498 and 583"
...

> s
  ...
       sentiment
1       Neutral
2       Negative
3       Positive
...
```

Finally, it is possible to calculate the total number of sentences for each of the six categories previously mentioned, using `calculate_total_presence_sentiment(x)`.

There is also the possibility of using a custom dictionary so that the package can also be used with texts written in a language different from English. In this case, the function used for calculating the polarity score, assigning a sentiment category, or counting the number of sentences in each category have to be substituted by the following one:

```
calculate_custom_option(x, positive_words, negative_words)
```

where *option* is equal to `score`, `sentiment`, and `total_presence_sentiment` for the three tasks, respectively, `positive_words` is a vector of positive terms, and `negative_words` is a vector of negative terms.

### *3.4* `sentimentr`

The `sentimentr` package (Rinker, 2019) calculates semantic polarity at a sentence level and optionally aggregates the score at a document level or according to a grouping variable.

The scoring function used in the package incorporates the weights attributed to *valence shifters*, including negators, amplifiers (intensifiers), de-amplifiers (downtoners), and adversative conjunctions (Polanyi and Zaenen, 2006). These shifters modify the sign and the intensity of the polarity in a text. For instance, negators or adversative conjunctions can reverse or overrule the entire semantic orientation of a sentence and hence, its sentiment. Since valence shifters occur frequently, simple dictionary lookup methods cannot be able to model the semantic orientation accurately. Packages that use dictionary lookups are faster and more effective when the goal is to analyze the general sentiment of a piece of literature. As an example, the methods offered by `syuzhet` are accurate but more prone to error in the presence of valence shifters in the texts.

The package imports texts as raw character vectors and segments them into sentences. Sentence boundary disambiguation is performed with the function `get_sentences`, a parser based on regular expressions. The main functions for calculating polarity scores are `sentiment` and `sentiment_by`. The first function allows users to easily alter (add, change, replace) the default term polarity with a valence shifters dictionary. The default command string is:

```
sentiment(x, polarity_dt=lexicon::hash_sentiment_jockers_rinker,
        valence_shifters_dt=lexicon::hash_valence_shifters,
                    n.before=5, n.after=2,...)
```

where `x` is a `get_sentences` object or a raw character vector, `polarity_dt` is a data table of positive/negative terms, `valence_shifters_dt` is a data table of valence shifters, `n.before` and `n.after` are the number of terms to consider as a window around the positive/negative terms in order to apply the valence shifters' weights. By default, a combination of an augmented version of the Jocker's dictionary (initially developed for the `syuzhet` package) and the Rinker's dictionary (an augmented version of the Hu and Liu dictionary) available in the `lexicon` package is used. The valence shifters' table, also recalled from `lexicon`, lists the shifters and considers a different integer key for negators (1), amplifiers (2), de-amplifiers (3), and adversative conjunctions (4). In the polarity function, it is also possible to change the default parameters related to the amplifier and adversative weights. The scores are measured on a con-tinuous scale on an interval of ]-∞;+∞[. A more detailed explanation of how this scoring function works can be found in Balbi, Misuraca, and Scepi (2018).

The results are reported in a dataframe where the columns represent the id number of the original vector passed to the function, the id number of the sentences within each text, the word count, and the polarity score, respectively.

```
1  library(sentimentr)
2  data(course_evaluations)
3  x<-course_evaluations$text
4  ## Calculating the sentiment scores of the text by sentence
5  s<-sentiment(x, polarity_dt = lexicon::hash_sentiment_jockers_rinker,
6     valence_shifters_dt = lexicon::hash_valence_shifters, n.before = 5,
7     n.after = 2)
8
```

```
> s
     element_id sentence_id word_count    sentiment
1:          1           1         12   0.02886751
2:          2           1          4  -0.25000000
3:          3           1         10   0.23717082
...
```

The `sentiment_by` function calculates the polarity scores at a document level or by considering a grouping variable(s). The scoring function is the same used in `sentiment`:

$$\text{sentiment\_by}(x, \text{by=NULL},$$
$$\text{averaging.function=sentimentr::average\_downweighted\_zero}, ...)$$

where x is a `sentimentr` object, `by` is a list of one or more grouping variables used to collapse the sentences/texts, `averaging.function` is the function used to calculate the polarity score on average on the entire text or group of texts. By default, the function uses the original sentence_id values and averages the polarity scores by down-weighting the zeros. This approach is useful when we do not want neutral sentences to have a strong influence on the general sentiment since a value of 0 is attributed to sentences without a polarization. Alternatives are the `average_weighted_mixed_sentiment` function, that up-weights the negative values and down-weights the zeros, and the standard `average_mean` function. The function returns a data frame where the columns are the id number of the original vector passed into the function, the id number of the sentences within each element, the (total) word count, the standard deviation, and the average polarity scores. If the texts contain only one sentence, the standard deviation is not calculated, and an NA tag is imputed instead.

```
1  library(sentimentr)
2  data(course_evaluations)
3  x<-course_evaluations$text
4  ## Obtaining the sentiment of the text by grouping variables
5  ## Note in this example the grouping variable is the element_id
6  s_by<-sentiment_by(x, by = NULL,
7        averaging.function = sentimentr::average_downweighted_zero)
8
```

```
> s_by
     element_id word_count        sd ave_sentiment
  1:          1         12        NA    0.02886751
  2:          2          4        NA   -0.25000000
  3:          3         10        NA    0.23717082
...
```

At the end of the process, it is possible to obtain a graphical representation of the results. If we want to plot the polarity scores at a sentence level, the command string is:

```
                       plot(object,
   transformation.function=syuzhet::get_trasformed_value, ...)
```

where `object` is a `sentimentr` object, and the `transformation.function` is used to smooth the polarity scores. By default, the `get_trasformed_value` function from `syuzhet` is used to obtain a smoothed polarity across the text, as in Figure 4. The *duration* plotted on the x-axis of the plot is equivalent to the narrative time used in `syuzhet`.

### 3.5 `SentimentAnalysis`

The `SentimentAnalysis` package (Feuerriegel & Pröllochs, 2019) provides a collection of tools for performing and comparing different methods for sentiment analysis. It is possible to refer to several lexicons, such as the `qdap` dictionaries (Goodrich, Kurkiewicz, & Rinker, 2019), the *Harvard General Inquirer* (Stone, Dunphry, Smith & Ogilvie, 1966), the *Loughran-McDonald's financial dictionary* (Loughran & McDonald, 2011), the *Henry's Financial* (Henry, 2008). Moreover, it is possible to customize the dictionary, choosing among several approaches.

The polarity scores are computed through the function `analyzeSentiment`. One of the main characteristics of this function and the whole package is its strict dependence on the `tm` library (Feinerer, Hornik, & Meyer, 2008; Feinerer & Hornik, 2018). This library is one of the most known for performing data preprocessing and management in a Text Mining framework.

The users can directly customize text pre-processing by modifying the parameters in the function. The default command line is:

```
analyzeSentiment(x, language="english", aggregate=NULL,
    rules=defaultSentimentRules(), removeStopwords=TRUE,
                    stemming=TRUE, ...)
```

where `x` is a vector of strings or a more complex text object generated by the `tm` package, such as a collection of documents in plain text or a document-term matrix, and `rules` is a list with the method for calculating the score and, optionally, the dictionary. Different methods are implemented, such as `ruleSentiment`, that calculates the ratio between positive and negative terms on the total number of terms. By default, stop-words are removed, and terms are stemmed according to stemming algorithms recalled from the `SnowballC` package.

The function returns a data frame with word counts and polarity scores from each dictionary implemented in the package.

```
1  library(sentimentr)
2  data(course_evaluations)
3  x<-course_evaluations$text
4  ## Calculating polarity scores in SentimentAnalysis
5  library(SentimentAnalysis)
6  s<-analyzeSentiment(x, language="english", aggregate=NULL,
7  rules=defaultSentimentRules(), removeStopwords=TRUE, stemming=TRUE)

> s
    WordCount  SentimentGI  NegativityGI  PositivityGI ...
1          6  -0.16666667   0.33333333    0.16666667
2          3   0.33333333   0.00000000    0.33333333
3          5   0.20000000   0.00000000    0.20000000
...
```

Since the polarity scores are measured on a continuous scale, it is possible to convert each score into a sentiment category by using the functions `convertToBinaryResponse` and `convertToDirection`. These two functions transform the score vector into a factor object by tagging each value with respect to its sign. The result of the first function is a factor with two levels indicating a positive orientation – also counting neutral texts – or a negative orientation. The second one produces a three-level factor, classifying texts in positive, neutral, or negative.

In order to visualize the sentiment evolution of the documents, it is possible to use `plotSentiment`. The graphical representation is similar to the plot showed in Figure 1.

The library also gives the possibility to compare the results of the scoring to an existing response variable. The response variable can be continuous or binary. The command line is:

```
compareToResponse(sentiment, response_variable)
```

where `sentiment` is a matrix containing the scores obtained from several rules for each document and `response_variable` the vector with true responses. The comparison can be graphically displayed through the function `plotSentimentResponse`, obtaining a scatterplot with a trend line in the form of a generalized additive model.

The dictionary used in the analyses can be customized with the function:

$$SentimentDictionary \textit{Option}()$$

where *Option* is equal to `Wordlist`, `Binary`, or `Weighted`. If `Wordlist` is selected, the argument of the function is a vector listing the terms to consider, where for `Binary` is necessary to use as argument two vectors with positive and negative terms, respectively. The `Weighted` option requires a vector of terms and a vector of corresponding scores.

Furthermore, the library implements a novel approach to extract terms that have a statistically significant impact on the response variable (Pröllochs, Feuerriegel, & Neumann, 2015), by using the `generateDictionary()` function. By default, the estimation is performed via *LASSO* regularization (Tang, Alelyani, & Liu, 2014), a statistical approach aiming at selecting the most relevant terms by an exogenous response variable. Moreover, it is possible to apply several estimation models such as elastic net, ordinary least squares, generalized linear model or spike-and-slab regression. By default, the command line is:

```
generateDictionary(DTM, response, language="english",
modelType="lasso", filterTerms=NULL, control=list(),
minWordLength=3, sparsity=0.9, weighting=function(x)
        tm::weightTfIdf(DTM, normalize = FALSE))
```

where `DTM` is a document-term matrix obtained from the library `tm`, `response` is a numeric vector containing the polarity scores assigned to the different documents.

Several functions allow exploring the generated dictionary. A simple overview can be displayed through the function `summary`.

```
1  library(sentiment)
2  data(course_evaluations)
3  x<-course_evaluations$text
4  ## Creating the document x term matrix
5  library(tm)
6  DTM<-DocumentTermMatrix(VCorpus(VectorSource(x)))
7  ## Generating dictionary with LASSO regularization
8  response<-course_evaluations$sentiment
9  library(SentimentAnalysis)
10 dict<-generateDictionary(DTM, response)
```

```
> dict
Type: weighted (words with individual scores)
Intercept: 0.2978264
-1.02 hated
-1.00 least
-0.88 own
-0.79 wasn't
-0.75 failed
...

> summary(dict)
Dictionary type:  weighted (words with individual scores)
Total entries:    299
Positive entries: 92 (30.77%)
Negative entries: 207 (69.23%)
Neutral entries:  0 (0%)

Details
Average score:      -0.09238678
Median:             -0.002648496
Min:                -1.023082
Max:                0.8005375
Standard deviation: 0.2814393
Skewness:           -0.5512306
```

A *kernel density estimation* (KDE) can also visualize the distribution of positive and negative terms (Figure 5). This type of plot allows inspecting the skewness of the dictionary's sentiment distribution, revealing if there are more positive terms than negative terms or vice versa.

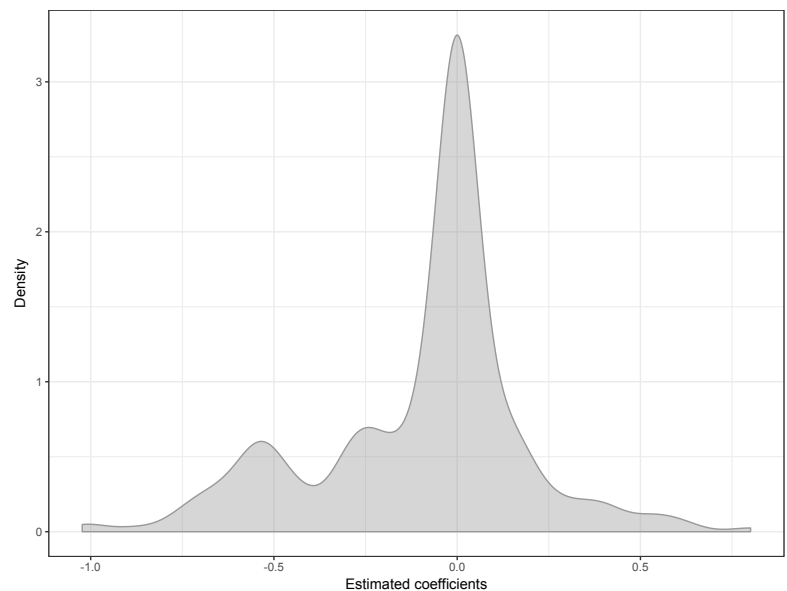

FIGURE 5. *Dictionary's sentiment distribution of the course_evaluation dataset in terms of KDE.*

It is possible to compare the generated dictionary to other dictionaries, regarding similarities and dissimilarities, via `compareDictionary`. The function computes several metrics, such as the overlap or the correlation between the dictionaries. It is important to note that the comparison works if both dictionaries are of the same type (wordlist, binary, or weighted).

`SentimentAnalysis` can be adapted to other languages different from English. In this case, it is necessary to change the parameters for document pre-processing (language, stopwords) and the dictionary used for polarity calculation.

### 3.6 `meanr`

The `meanr` package (Schmidt, 2019) provides a simple method for computing the polarity score of a set of documents. Each term is scored as positive or negative according to a fixed dictionary, the lexicon developed by Hu and Liu. Differently from the other packages, it does not require any additional `R` libraries to work. The only function implemented is:

```
score(x, nthreads=meanr.nthreads())
```

where `x` is a character vector of strings and `nthreads` represents the number of threads to use. The threads are by default the total number of cores and hyperthreads on the user system.

The scoring function operates at a document-level, using for multiple documents a parallel computing approach via OpenMP. Since the function removes the punctuation and converts all the characters belonging to the strings into lowercase, a pre-processing step is mostly unnecessary. It estimates the semantic orientation by checking if each term is positive, negative or neutral. According to Hu and Liu's lexicon, the function assigns a value +1 to each positive term and -1 to each negative term. A term is assumed to be neutral when it is not checked as positive or negative, assigning a null value. The polarity score is calculated as the difference between the count of positive and negative terms. The output for each string is the count of positive and negative terms, the polarity score, and the total number of terms. It is neither possible to customize the dictionary nor to use a different weighting scheme to calculate the scores.

```
1  library(sentimentr)
2  data(course_evaluations)
3  x<-course_evaluations$text
4  ## Calculating sentiment scores in meanr
5  library(meanr)
6  s<-score(x)

> s
  positive negative score wc
1        1        1     0 14
```

```
2       1       0      1  6
3       1       0      1 13
...
```

### 3.7 Other packages for Sentiment Analysis

There are other `R` packages for SA, which are not discussed in this review. The package `sentometrics` (Borms, *et al.*, 2019) offers an integrated framework for textual sentiment time series aggregation, allowing a quantification of the semantic polarity and considering its time distribution. In this perspective, it is possible to perform predictive analyses on the sentiment temporal evolution. Other packages are not specifically designed for SA but provide functions for calculating the polarity scores. As an example, `tidytext` (De Queiroz, *et al.*, 2019) contains a `get_sentiments` function that calculates the polarity scores of a document collection through the same lexicons used by `syuzhet`, in a more comprehensive Text Mining framework based on the tidy-data normalization scheme (Wickham, 2014).

## 4. Discussion

After describing the main features of the `R` packages designed to perform SA, here we present and discuss the results of a comparative study by considering the different viewpoints.

Firstly, we focused our attention on how polarity scores are calculated in the packages. All the functions start with the same step: tagging the polarized terms in each sentence according to a reference lexicon. In `syuzhet`, `Rsentiment`, and `meanr` the polarity of a text is simply calculated as the algebraic sum of the term scores. `SentimentAnalysis` also calculates polarity scores in a straightforward way, considering the ratio between positive and negative terms on the total number of terms. A more complicated but more effective way of calculating polarity scores is implemented in `sentimentr`. As we said above, the scoring function used in the package incorporates the weights attributed to valence shifters, which can modify the sign and the intensity of the text polarity. In the presence of valence shifters, simple dictionary lookup methods are not able to accurately model the semantic orientation. On the other hand, if the aim is only to analyze the general sentiment of a text, dictionary lookup methods are faster and more effective. Another aspect to consider is that `sentimentr` performs the polarity detection at a sentence-level, then averaging the different scores allows to calculate the semantic orientation at a text-level. The scoring function of `Rsentiment` also checks for the POS of each term, considering the presence of negations and/or amplifiers, but the mathematical computation of the scores is simpler than the one offered by `sentimentr`.

Another aspect to consider is that polarity scores are affected by the choice of the reference lexicon. Therefore, it becomes very important having the possibility of calculating the scores by applying the same approach with different lists of polarized terms. Table 2 reports the different types of lexicon that can be used in the packages. We noted that all the packages, except for `meanr`, provide the possibility of calculating the polarity scores with a customized dictionary. In this way, it is not only possible to choose among the different lexicons proposed in literature, but also to create a dictionary based on proper research needs. More in-depth, only `sentimentr`, `SentimentAnalysis`, and `syuzhet` allow attributing different weights to polarized terms by considering not only the sign of the polarity but also a number that reflects its intensity. Finally, it is worth noting that only `syuzhet` includes a lexicon for the analysis of multilingual collections (*nrc*), but it is also true that a multilingual lexicon can be imported as custom dictionary in the packages providing this option.

TABLE 2.
*Dictionary selection in the analyzed SA packages*

| Packages | Custom Dictionary | Weighted Dictionary | Multilingual |
|---|---|---|---|
| `meanr` | − | − | − |
| `RSentiment` | + | − | − |
| `sentimentr` | + | + | − |
| `SentimentAnalysis` | + | + | − |
| `syuzhet` | + | + | + |

The real peculiarity of `syuzhet` is the possibility to obtain an emotional categorization of texts by applying the function `get_nrc_sentiment()`.

For a more interesting comparison among the packages, in the following, we performed a performance and an accuracy analysis using as benchmark the *course_evaluation* dataset. It is important to consider that each function was run with the default options, hence the dictionaries used to calculate the polarity scores were different.

### 4.1 Performance analysis

To compare the different packages, we carried on a performance analysis on the different functions used to calculate the polarity scores of the case example. Tests were performed on a MacBook Pro 2.9 GHz Intel Core i9, 32 GB 2400 MHz DDR4 with macOS Mojave ver. 10.14.5, by using `R` 3.5.3 "Great Truth" through the IDE RStudio Desktop.

TABLE 3.
*Running times of the polarity scores functions (in seconds)*

| Packages | Function | User | System | Elapsed |
|---|---|---|---|---|
| `meanr` | `score()` | 0.004 | 0.005 | 0.003 |
| `syuzhet` | `get_sentiment()` | 0.175 | 0.034 | 0.210 |
| `sentimentr` | `sentiment_by()` | 1.070 | 0.004 | 0.249 |

| SentimentAnalysis | analyzeSentiment() | 1.518 | 0.053 | 1.601 |
| Rsentiment | calculate_score() | 78.166 | 3.047 | 67.307 |

For each package, the running time was obtained by introducing the `system.time` function in the codes showed above. This function calculates the time execution of an `R` expression, returning a vector of five values, the *user time* (time spent executing the code), the *system time* (time spent executing system functions called by the code), the *elapsed time* ("real" time since the process started), and the *user* and *system time* for any child process. The user time gives the CPU time spent by the specific process (i.e., the `R` session), while the system time gives the CPU time spent by the operating system on behalf of the current process. The elapsed time is the actual time taken from the start to the end of the process so that it can be affected by other processes running meanwhile on the computer. In Table 3, the running times for each function are reported.

Looking at the user time, we noted that the fastest function was `score()` from `meanr` package. This function operates at a text-level, using for multiple documents a parallel computing approach via OpenMP. Moreover, the polarity score is calculated as the difference between the count of positive and negative terms, without any other computation. The second function, in terms of speed, was `get_sentiment()` from `syuzhet` with 0.175 seconds, followed by `sentiment_by()` from `sentimentr` with 1.070 seconds, and `analyzeSentiment()` from `SentimentAnalysis` with 1.518 seconds. The slowest function was `calculate_score()` from `Rsentiment`, which took more than one minute to calculate the different polarity scores. This time probably depends on the on-screen prints of each processed text, making the process slower than the other ones. It is necessary to consider that the different results were obtained for each function by using the default options. Moreover, these results depend on the technical characteristics of the computer used to perform the analyses.

### 4.2 Accuracy analysis

To evaluate how the different `R` packages calculate the polarity scores, we analyzed in-depth the results obtained above. The *course_evaluations* dataset by Welch & Mihalcea contains the student comments as well as the true sentiment classification. The authors manually labeled the sentiment toward each comment. When no explicit sentiment was expressed or evident, it was assumed to be neutral. In the analyzed subset, we found that 53% of sentences was positive, 38.34% negative, and 8.66% neutral, with a score of +1, -1, and 0, respectively.

Each function used to calculate the polarity returns values on different scales. In `meanr` and `Rsentiment`, scores are measured on a discrete scale. The polarity scores calculated by `syuzhet`, `sentimentr`, and `SentimentAnalysis` are instead measured on a continuous scale. Both discrete and continuous scales are unbounded. To perform the comparison, we converted each score into a sentiment category – labelled as Neg, Neu, and Pos – by considering only the sign of the score. Since `Rsentiment` considers interrogative sentences as sarcastic and labels them with a score of 99, we also marked these sentences as neutral.

Table 4 reports for each package the sentiment category distribution of the comments, together with the corresponding misclassification rate. Among the different packages, `sentimentr` reproduced more faithfully the true classification of the comments, with only a 24.20% of misclassified. It is interesting to note that all the functions overestimated positive comments.

TABLE 4.
*Comparison of the true classification with the obtained results*

|  | Neg (%) | Neu (%) | Pos (%) | Misclassified (%) |
|---|---|---|---|---|
| true classification | 38.34 | 8.66 | 53.00 | – |
| `meanr` | 16.61 | 21.91 | 61.48 | 37.28 |
| `syuzhet` | 17.14 | 7.24 | 75.62 | 32.86 |
| `sentimentr` | 33.57 | 6.01 | 60.42 | 24.20 |
| `SentimentAnalysis` | 15.72 | 22.61 | 61.66 | 44.34 |
| `Rsentiment` | 22.97 | 19.96 | 57.07 | 31.09 |

To better evaluate the effectiveness of the different functions, we also graphically analyzed the results. Figure 6 displays the five confusion matrices, obtained by cross-tabulating for each package the true classification and the predicted categories. Starting from each confusion matrix, we calculated the primary indices used in the framework of supervised classification (Sokolova & Lapalme, 2009). These validation measures consider how much the predicted categories correspond to the true classification, reflecting different viewpoints.

The *accuracy* is the overall fraction of comments correctly classified, the *precision* is the agreement of the true labels of the comments with those obtained by the given function, the *recall* is the effectiveness of the function to identify the true labels, the *F-measure* is the harmonic mean of precision and recall representing the relation between true labels and those given by the function. In a multi-class categorization, as in sentiment analysis, the average per-class precision ($P_M$), recall ($R_M$), and F-measure ($F1_M$) are usually calculated. This approach to validation is referred to as *macro-averaging*. In Table 5, the measures calculated on each package are showed.

TABLE 5.
*Validation measures for each package (macro-averaging)*

| Package | Ov. Acc. | $P_M$ | $R_M$ | $F1_M$ |
|---|---|---|---|---|
| `meanr` | 0.627 | 0.586 | 0.565 | 0.521 |
| `syuzhet` | 0.671 | 0.605 | 0.526 | 0.526 |
| `sentimentr` | 0.758 | 0.649 | 0.614 | 0.626 |
| `SentimentAnalysis` | 0.557 | 0.506 | 0.479 | 0.445 |
| Rsentiment | 0.689 | 0.614 | 0.621 | 0.585 |

All the reported values confirm that the function used in `sentimentr` to calculate the polarity scores outperforms the others in recognizing the true polarity of the comments. This result depends on the use of valence shifters, which emphasize the negative or the positive term polarity, as well as the text-level approach to polarity detection.

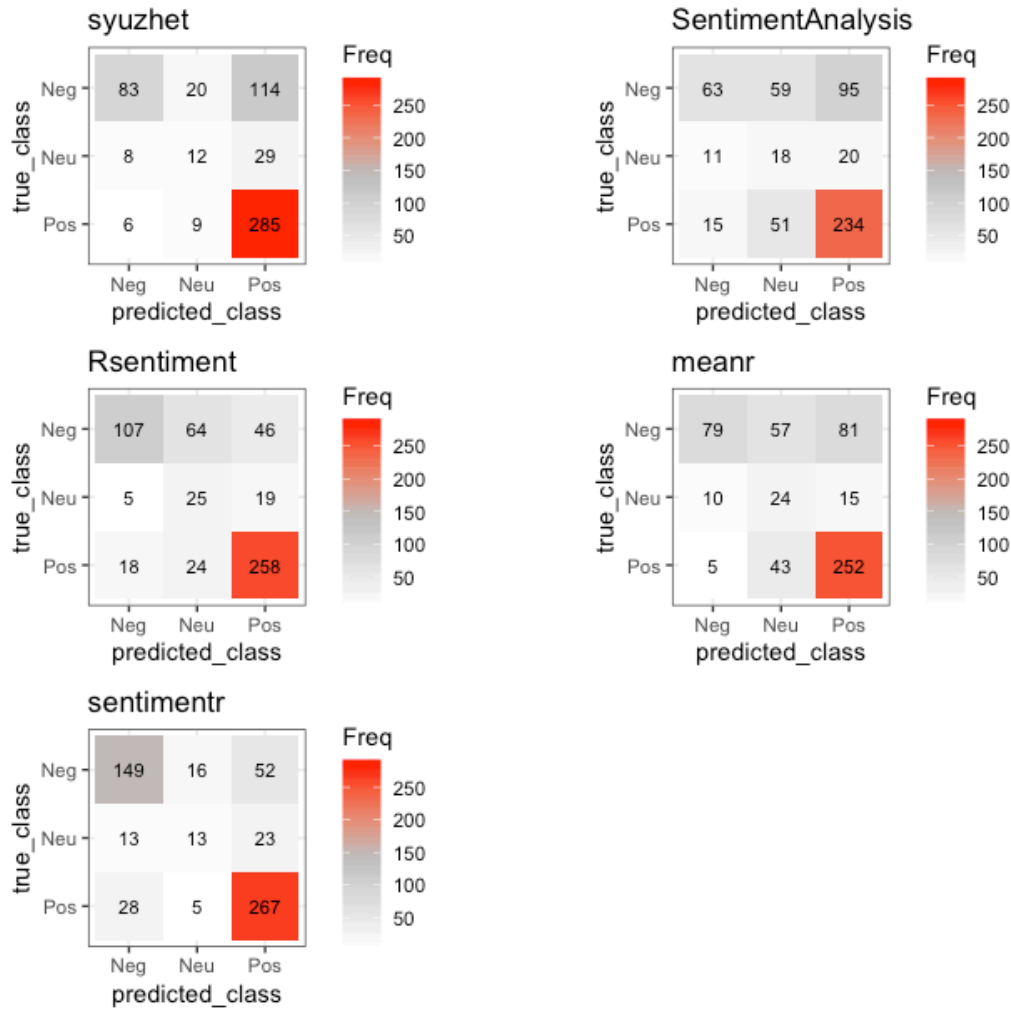

FIGURE 6. *Confusion matrices of the five sentiment classifications produced by the analyzed packages*

## 5. Conclusions

The aim of the review presented in this paper is to compare the R packages designed for SA, considering their functionality and ease of use. We think that this work can be useful for new users who are familiar with the R environment, but not with SA applications, especially in the framework of Education. A focus on student comments that consider the semantic orientation of the different opinions can significantly help the evaluation of learning activities. SA, as dis-cussed in the paper, can be considered an exploratory analysis. However, calculating the polarity scores can be seen as just the first step of more sophisticated supervised classification processes, e.g., when the aim is to categorize new texts considering the knowledge base that emerges from a training step on labelled texts.

The analysis of the *course_evaluations* dataset showed that `meanr` is the fastest package for calculating polarity scores, but it produces the less accurate results. Better results in terms of accuracy can be obtained by using `syuzhet` and `Rsentiment`, with the first package faster than the second one. The highest level of accuracy can be reached by calculating polarity scores with `sentimentr`, which was the third package for running times in our case study. The high level of accuracy depends on the use of valence shifters. Moreover, `sentimentr` had also the advantage of calculating polarity scores both at a sentence-level and at the document-level. Regardless of the different methods for calculating polarity scores, another aspect to consider is the possibility of importing and customizing the dictionary of polarized terms. Except for `meanr`, all the presented packages provide this option with more or less a similar difficulty of implementation.

It is important to consider that the use of different lexicons can affect the results of the analyses. Our choice of computing polarity scores for each function by using the default options, therefore with different lexicons, can be considered a limit of this study. Nevertheless, we supposed that the authors of the packages suggest the default options as the best choice for calculating polarities with respect to the implemented method. It is left to the readers and the practitioners the choice of the best solution to cope with their specific informative needs.

### Declaration of Conflicting Interests

The authors declared no potential conflicts of interest with respect to the research, authorship, and/or publication of this article.

### Funding

The authors received no financial support for the research, authorship, and/or publication of this article.